\documentclass[aps,prb,superscriptaddress,reprint,showpacs,twocolumn,preprintnumbers,amsmath,amssymb,floatfix]{revtex4-1}
\usepackage{graphicx}
\usepackage{color}
\usepackage[latin1]{inputenc}
\usepackage{amssymb}
\usepackage{bm}
\usepackage{stmaryrd}
\usepackage{wasysym}
\usepackage{subfigure}
\usepackage{afterpage}

\newcommand{\csfe}{CsFe$_2$As$_2$ }
\newcommand{\rbfe}{RbFe$_2$As$_2$ }

\begin{document}

\title{Charge and nematic orders  in AFe$_2$As$_2$ (A= Rb,Cs) superconductors}

\author{M. Moroni}
\affiliation{Department of Physics, University of Pavia-CNISM,
I-27100 Pavia, Italy}
\author{G. Prando}
\affiliation{Department of Physics, University of Pavia-CNISM,
I-27100 Pavia, Italy}
\author{S. Aswartham}
\affiliation{IFW Dresden, Helmholtzstrasse 20, 01069 Dresden,
Germany}
\author{I. Morozov}
\affiliation{IFW Dresden, Helmholtzstrasse 20, 01069 Dresden,
Germany}
\affiliation{Lomonosov Moscow State University, 119991 Moscow, Russia
}
\author{Z. Bukowski}
\affiliation{Institute of Low Temperature and Structure Research, Polish Academy of Sciences, 50-422 Wroclaw, Poland}
\author{B. B\"uchner}
\affiliation{IFW Dresden, Helmholtzstrasse 20, 01069 Dresden,
Germany}
\affiliation{ Institut f\"ur Festk\"orperphysik, Technische Universit\"at Dresden, D-01171 Dresden,
Germany}
\author{H.J. Grafe}
\affiliation{IFW Dresden, Helmholtzstrasse 20, 01069 Dresden,
Germany}
\author{P. Carretta}
\affiliation{Department of Physics, University of Pavia-CNISM,
I-27100 Pavia, Italy} \email{pietro.carretta@unipv.it}

\begin{abstract}
 We discuss the results of $^{75}$As Nuclear Quadrupole Resonance (NQR) and muon spin
relaxation measurements in AFe$_2$As$_2$ (A= Cs, Rb) iron-based
superconductors. We demonstrate that the crossover
detected in the nuclear spin-lattice relaxation rate $1/T_1$ (around 150 K in \rbfe and around 75 K in \csfe), from
a high temperature nearly localized to a low temperature
delocalized behaviour, is associated with the onset of an
inhomogeneous local charge distribution causing the broadening or
even the splitting of the NQR spectra as well as an increase in
the muon spin relaxation rate. We argue that this crossover, occurring at temperatures well above the phase transition to the nematic long-range order, is associated with a charge disproportionation at the Fe sites
induced by competing Hund's and Coulomb couplings. In \rbfe around 35 K, far below that crossover temperature, we observe a peak in the NQR $1/T_1$ which is possibly associated with the critical slowing down of electronic nematic fluctuations on approaching the transition to the nematic long-range order.
\end{abstract}


\maketitle


\section{\label{sec:intro} Introduction}

The high temperature superconducting cuprates
are paradigmatic examples of strongly correlated electron systems close to a Mott-Hubbard transition,\cite{MHub} where the large electron Coulomb repulsion $U$
competes with the tendency of the electrons to delocalize. Electronic correlations remain sizeable even at hole doping levels leading to high temperature superconductivity and the competition between different energy scales gives rise to several crossovers and phases, the most debated one being the charge density wave (CDW), which progressively fades away as the
doping increases \cite{Marc,Tranquada,Ghiringhelli,Hucker}. The effect of electronic correlations
in iron-based superconductors (IBS)\cite{IBS} is more subtle. At
variance with the cuprates, in IBS  the Fermi level typically crosses five bands associated with the Fe 3$d$
orbitals, leading to a rich phenomenology in the normal as well as
in the superconducting state.\cite{bands,Johnston} The most relevant phenomena emerging from the electronic structure of the IBS are the electronic nematic order\cite{nema1}, developing over a wide portion of IBS phase diagrams, and the presence of an orbital selective behaviour,\cite{orbsel1} which can eventually lead to orbital selective Mott transitions.\cite{orbselMott}

While the nematic fluctuations are detected over a broad charge doping range,\cite{nemNMR,Gallais} the orbital selective behaviour becomes relevant only upon approaching half filling (5 e$^-$/Fe), where the effect of electronic correlations gets more sizeable.\cite{orbsel1} In particular, in the hole-doped AFe$_2$As$_2$ (A=Cs, K or Rb) IBS, with 5.5 e$^-$/Fe, it has been theoretically pointed out since almost 10 years ago\cite{orbsel1} that while electrons in d$_{xy}$ bands are close to localization, the other bands remain metallic with much lower effective masses. This phenomenology, driven by the Hund's coupling $J$, which tends to decouple charge excitations in the different orbitals, has so far been evidenced by many experiments.\cite{orbselexp1,mass1,mass2} More interestingly, the proximity to localization for certain bands and the competition between $J$ and $U$ could eventually lead to a inhomogeneous charge distribution among the Fe ions.\cite{Isidori} Namely to a charge order, as in the cuprates, even if driven by a different mechanism. In fact, the large Hund's
coupling $J$ promotes the electron transfer to generate high spin (HS) configurations and accordingly to a modulation of the charge distribution on the Fe ions.\cite{Isidori}


In this manuscript we report on $^{75}$As nuclear quadrupolar resonance (NQR) measurements in \rbfe and \csfe single crystals, complementing our previous results on \rbfe powders.\cite{civardi} NQR has the great advantage of allowing to probe the equilibrium properties of the system without needing to perturb its electronic state and the lattice by applying an external magnetic field, strain or by inducing electronic transitions. As for the powders, we found also in the crystals a broadening of the NQR line below a characteristic temperature $T^*$, the same temperature where a crossover in $^{75}$As $1/T_1$ to a low temperature power law behaviour is detected. The observed NQR line broadening cannot be associated with a static nematic order but rather signals the onset of a charge order, possibly described within the framework of the charge disproportionation model. The changes in the electronic structure taking place at T$^*$ are observed to affect also the muon spin relaxation rate. Finally, we show that $^{75}$As NQR $1/T_1$ shows a clear anomaly at T$_{nem}\simeq 35$ K, evidencing the slowing down of nematic fluctuations on approaching the Ising nematic order recently detected by elastoresistivity measurements.\cite{XYnematic}

\begin{figure}[h!]
\vspace{10.4cm} \includegraphics{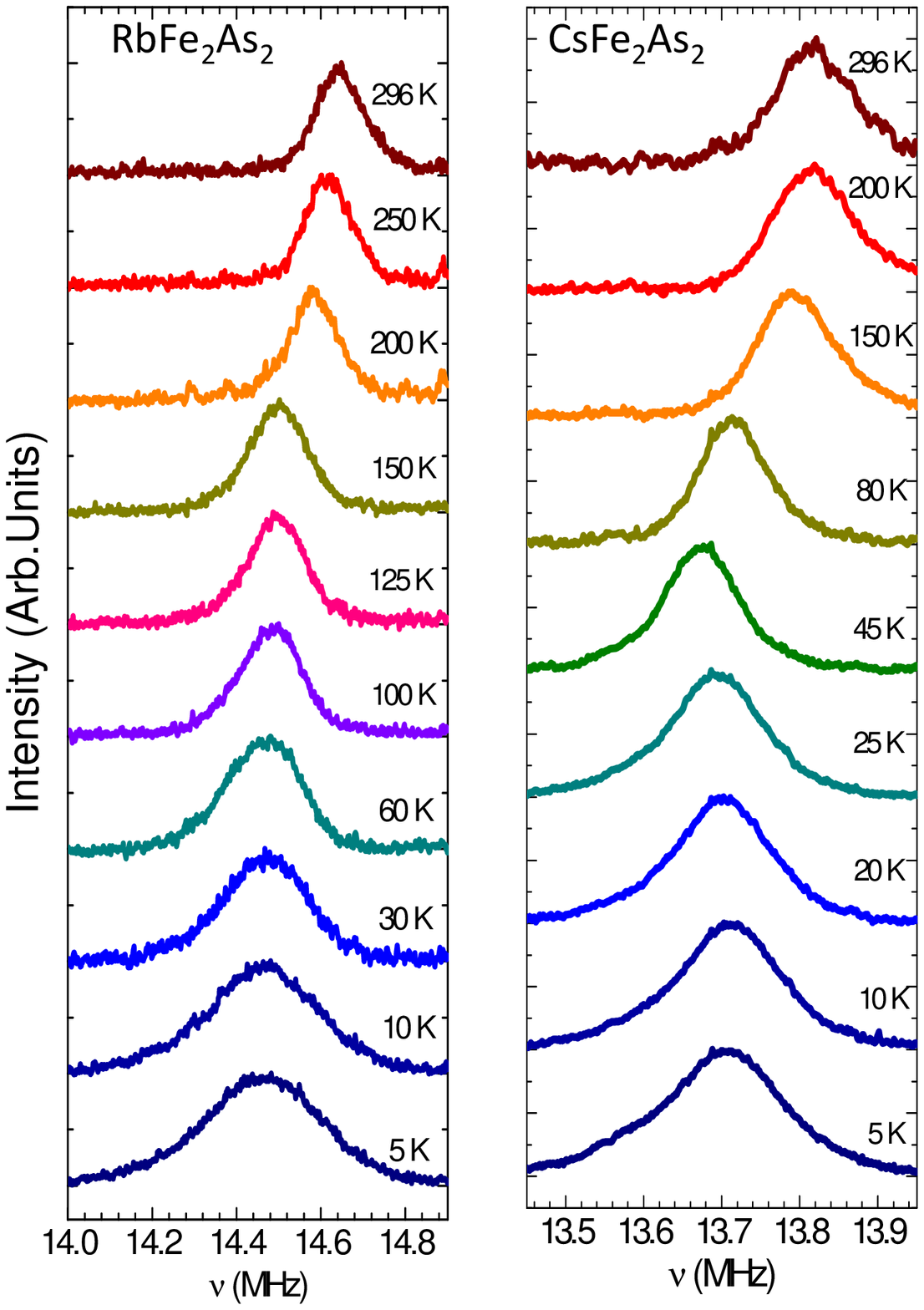} \caption{ (Color online)
Temperature evolution of $^{75}$As NQR spectra in \rbfe (left) and
\csfe (right) single crystals. } \label{NQRspectra}
\end{figure}

\section{\label{sec:methods} Experimental methods and results}

The measurements presented in this work were performed both on
powders, for \rbfe, and on single crystals of (Cs,Rb)Fe$_2$As$_2$. The powders
were grown following the procedure described in Ref.\onlinecite{Bukowski}
while the single crystals were grown by use of the self-flux
technique, as described in Ref.\onlinecite{xtal}. Owing to their air
sensitivity care was taken to avoid exposure to air. 

\begin{figure}[h!]
\vspace{11cm} \includegraphics{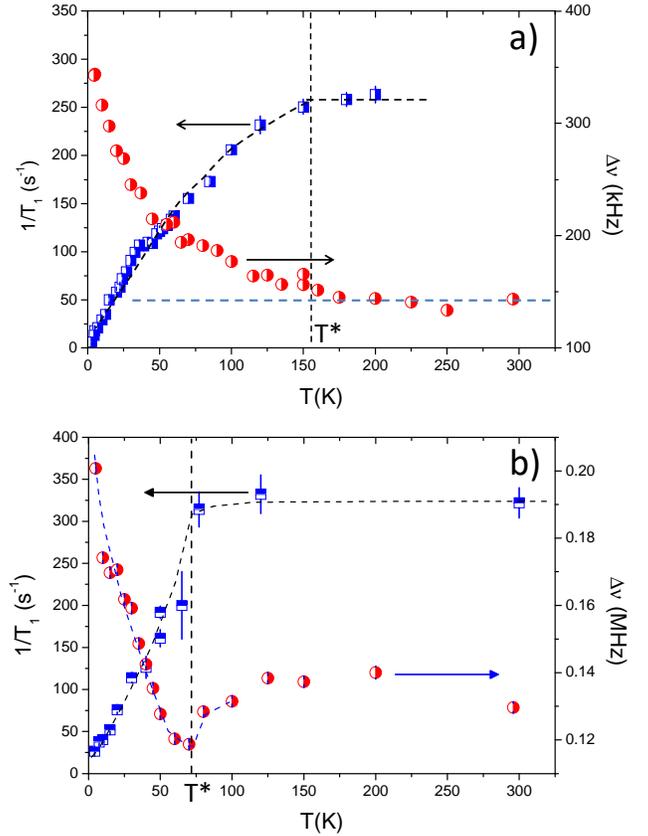} \caption{ (Color online) Temperature
dependence of $^{75}$As NQR nuclear spin-lattice relaxation rate
$1/T_1$ (blue squares) and linewidth $\Delta\nu$ (red circles) for \rbfe (a) and for \csfe (b) single crystals.
The vertical dashed line evidences $T^*$. The other dahed lines
are guides to the eye. } \label{FigT1nuQ}
\end{figure}

\begin{figure}[h!]
\vspace{9.5cm} \includegraphics{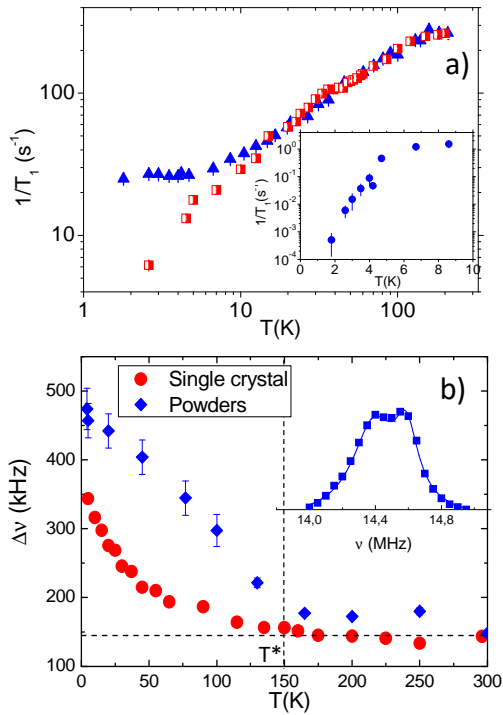} \caption{ (Color online) a) Comparison of
the temperature dependence of $^{75}$As NQR $1/T_1$ in \rbfe powders (blue triangles) and in \rbfe crystals (red squares). For the powders, we report in the main panel the data for the fast relaxing component and in the inset the data for the slow relaxing component emerging below 10 K (data from Ref.\onlinecite{civardi}. b) Comparison of the full width at half maximum of
$^{75}$As NQR spectra in \rbfe single crystals and powders (data from Ref.\onlinecite{civardi}). In
both cases an increase of the linewidth is noticed below
$T^*\simeq 150$ K. The dashed horizontal line helps to visualize
the high temperature linewidth in the single crystals. In the inset the $^{75}$As NQR spectrum in \rbfe powders, at $T= 5$ K, is shown.\cite{civardi}}
\label{DnuRb}
\end{figure}

$^{75}$As NQR spectra were obtained by
recording the amplitude of the Fourier transform of half of the
echo, detected after a $\pi/2-\tau-\pi$ pulse sequence, as a
function of the irradiation frequency. The delay $\tau$ was kept much
lower than the transverse decay time $T_2$ \cite{civardi} in order to avoid an unwanted reduction of the echo signal intensity. The NQR spectra for
the \csfe and \rbfe single crystals are reported in
Fig.\ref{NQRspectra}. One observes in both cases a broadening of
the NQR line below a temperature $T^*\simeq 150$ K for \rbfe and
$T^*\simeq 75$ K for \csfe (see Fig. \ref{FigT1nuQ}).\cite{Tstar}  In the \rbfe
powder sample one observes a more pronounced broadening with
respect to \rbfe single crystal (see Fig. \ref{DnuRb}) and, as it was pointed out by Civardi et al. \cite{civardi}, at low
temperature two bumps in the
NQR spectrum are detected, which indicates a modulation of the local charge
distribution across the lattice.

\begin{figure}[h!]
\vspace{6cm} \includegraphics{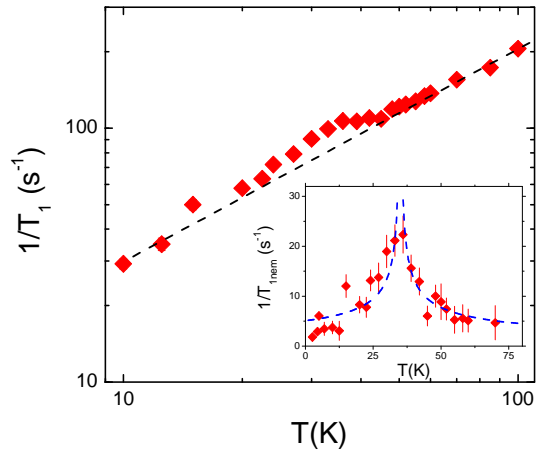} \caption{ (Color online) Temperature
dependence of $^{75}$As NQR nuclear spin-lattice relaxation rate
$1/T_1$ in \rbfe single crystal, in the 10-100 K temperature range. The dashed line is the power law behaviour found for $T\leq 25$ K and for 45 K$\leq T\leq$ T$^*$. In the inset the same data are shown after subtracting the power law behaviour $a\cdot T^b + c$, with $a=4.7 $ K$^{-b}$s$^{-1}$,$b=0.82$ and $c= 6$ s$^{-1}$. The dashed lines in the inset show a critical behaviour of the form $1/T_{1nem}\propto [\pm (T-T_{nem})]^{-1/2}$, with T$_{nem}=35$ K.} \label{FigT1nem}
\end{figure}

$^{75}$As nuclear spin-lattice relaxation rate $1/T_1$ was
measured by recording the recovery of the nuclear magnetization
$m(\tau)$ after a saturation recovery pulse sequence. In the
single crystals the recovery of the nuclear magnetization could be
fit with a single component\cite{suppl}
\begin{equation}
m(\tau)=m_0[1-f\, \cdot e^{-(3\tau/T_1)^{\beta}}]\;\mbox{,}
\label{eqrecovery}
\end{equation}
with $f\simeq 1$, a factor accounting for a non-perfect saturation, and $\beta$ a stretching exponent close to unity. This is different from the low temperature results for \rbfe
powder samples where two exponential components were
detected below 15 K\cite{civardi}: a slow relaxing one and a fast relaxing
one. The temperature dependence of $^{75}$As NQR $1/T_1$ is shown in
Fig.\ref{FigT1nuQ}. While above $T^*$ $^{75}$As $1/T_1$ is almost
constant, below $T^*$ one observes a power law
decrease, in agreement with previous reports.\cite{Khune,civardi,Wu} The behaviour in the \rbfe single crystals and powders is quite similar down to about 10 K (see Fig.\ref{DnuRb}). Below that
temperature the spin-lattice relaxation rate measured in the
powder, associated with the fast component tends to flatten,
while the $1/T_1$ for the slow component decreases faster.\cite{civardi} Namely,
below 10 K the behaviour found in the single crystals is in
between that found for the two components in the powders. As it will be emphasized in the following, in the \rbfe single crystal a clear bump in $1/T_1$ temperature dependence emerges around 35 K (see Fig.\ref{FigT1nem}). On the other hand, in \csfe we do not observe that anomaly.

The changes occurring at $T^*$ seem to affect also the zero field
muon spin relaxation (ZF$\mu$SR). ZF$\mu$SR measurements on \rbfe
powders were performed at ISIS facility on EMU beam line, while the ones on \rbfe and \csfe crystals were carried out at the Paul Scherrer Institute on GPS beam line. The powder sample was mounted in an air tight sample holder inside a glove
box in order to avoid exposure to air. The crystals were also mounted on a sample holder inside a glove box. They were first wrapped
with a 20 $\mu m$ silver foil, then sealed inside a 100 $\mu m$
kapton adhesive tape, mounted on the sample holder and then
transferred in an air tight bottle to the beam line. Once at the
beam line the sample holder was exposed to air only for the 
time needed to mount it on the stick and insert in the cryostat (about 2-3 minutes). Notice that during that time the crystal was in
anyway protected from air exposure by the kapton tape. The
crystals were mounted with the $c$ axis along the initial muon
polarization. A mosaic of 2 and 3 platelike crystals were used to cover a sufficiently large surface (about 1 cm$^2$). Since their thickness was of a few hundreds of microns we used kapton foil degraders to increase the muon stopping fraction inside the sample. 

In all cases the decay of the muon polarization
could be fit with a stretched exponential behaviour\cite{suppl}
\begin{equation}
P_{\mu}(t)= A_1 e^{-(\lambda t)^{\beta_1}}+ A_{bck}\,\,\, ,
\label{eqasy}
\end{equation}
with a temperature dependent exponent $\beta_1$. $\lambda$ is the
ZF$\mu$SR relaxation rate, $A_1$ the initial asymmetry and $A_{bck}$ a constant background due to the sample environment, mainly due to silver for the measurements on \rbfe powders. We point out that during the measurements at GPS, the initial asymmetry was reduced due to muonium formation in the kapton foil degraders.\cite{suppl}  

\begin{figure}[h!]
\vspace{14.2cm} \includegraphics{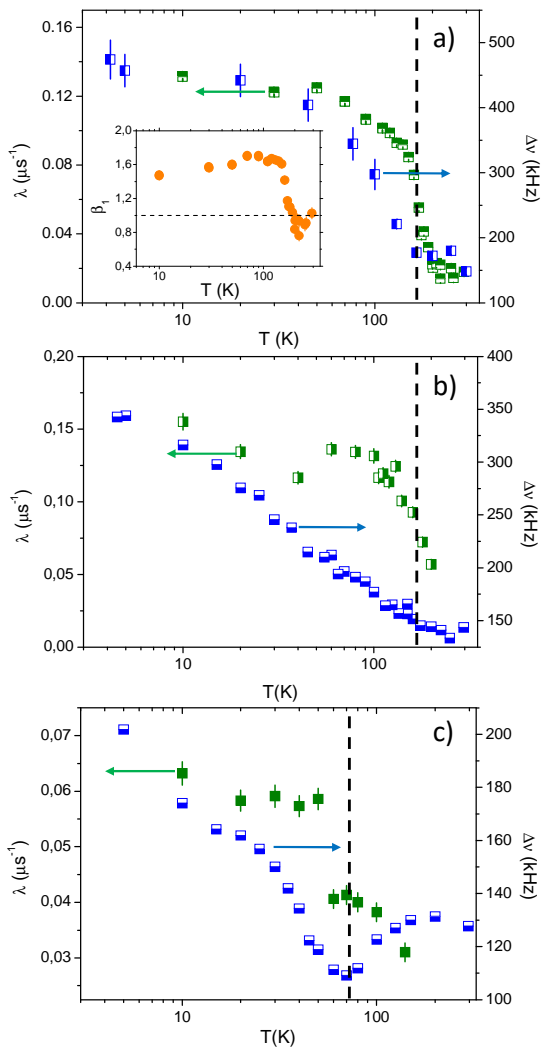} \caption{ (Color online) Temperature
dependence of the ZF$\mu$SR relaxation rate $\lambda$ and
$^{75}$As NQR linewidth $\Delta\nu$ for \rbfe powders (a) and
single crystal (b) and for \csfe (c). The vertical dashed line
evidences $T^*$. The other dashed lines are guides to the eye. The
inset in the figure at the top shows the temperature evolution of
the $\beta_1$ parameter in Eq.\ref{eqasy} } \label{ZFmuSR}
\end{figure}

The temperature dependence of $\lambda$ in \rbfe powders and in \rbfe and \csfe
crystals is shown in Fig. \ref{ZFmuSR}. In the \rbfe powders and
crystals there is a clear increase in the relaxation on cooling at
a temperature slightly higher than $T^*$ while in \csfe crystals
the increase is more modest and seems to take place slightly below $T^*$. It is
interesting to notice that in the powders when $\lambda$ increases
there is also a neat increase of $\beta_1$ (Fig.4), from values
slightly below 1 to about 1.8.

\section{\label{sec:discussion} Discussion}

For a nuclear spin $I=3/2$, as it is the case of $^{75}$As, the NQR
spectrum is characterized by a single line, associated with the
$m_I=\pm 1/2\rightarrow \pm 3/2$ transition, at a
frequency\cite{Abragam}
\begin{equation} \label{nqrspe}
\nu_Q= \frac{eQV_{ZZ}}{2h} \biggl(1+ \frac{\eta^2}{3}\biggr)^{1/2}
\,\,\, ,
\end{equation}
with $Q$ the nuclear electric quadrupole moment, $V_{ZZ}$ the main
component of the electric field gradient (EFG) tensor and $\eta$
its asymmetry $\eta= (V_{XX}-V_{YY})/V_{ZZ}$. Hence the NQR
spectrum probes the EFG at the nuclei induced by the local charge
distribution. As we have already pointed out,\cite{civardi}
although DFT calculations provide a reasonable estimate of the NQR
frequency they cannot account for many details of the physics of
these systems as the NQR line broadening. The broadening of the NQR line
indicates a modulation of the local charge distribution across the
lattice, as it is typically observed in the cuprates when a charge order develops.\cite{WuQ}  The different magnitude of the broadening observed in the \rbfe crystal and in the powders could tentatively be associated with the higher degree of disorder in the latter ones.  

The experimental observation that
both in \csfe and in \rbfe the broadening of the NQR spectrum
occurs at the same temperature $T^*$ where $1/T_1$ starts to
drop, suggests that at $T^*$ a change in the electronic
configuration is taking place. In particular, while the nearly
temperature independent behaviour of $1/T_1$ at high temperature
is typical of a strongly correlated electron system with nearly
localized electrons,\cite{NMRT1} the power law behaviour observed below $T^*$
indicates a more metallic behaviour. This phenomenology is
typically observed in Heavy Fermion (HF) compounds, as CeCu$_6$ for example,\cite{HFT1}  with $T^*$
corresponding to the coherence temperature below which the heavy
$f$ electrons delocalize into the Fermi sea. In \csfe and \rbfe
orbital selectivity gives rise to $3d$ bands with significantly
different effective masses\cite{orbsel1} and accordingly a phenomenology of
$1/T_1$ similar to that observed in HF can be expected. However,
there is an important difference: in HF compounds usually there is
not a change in the NQR spectra across the coherence temperature,
except when charge fluctuations are involved.\cite{HFcharge}
Here we observe a broadening of the spectra in the powders and crystals and, as it was remarked in Ref.\onlinecite{civardi}, the two bumps which emerge
in \rbfe powders at low temperature indicate the
breaking of the lattice translational symmetry, namely that there are at
least two inequivalent charge configurations around $^{75}$As
nuclei.

\begin{figure}[h!]
\vspace{6.3cm} \includegraphics{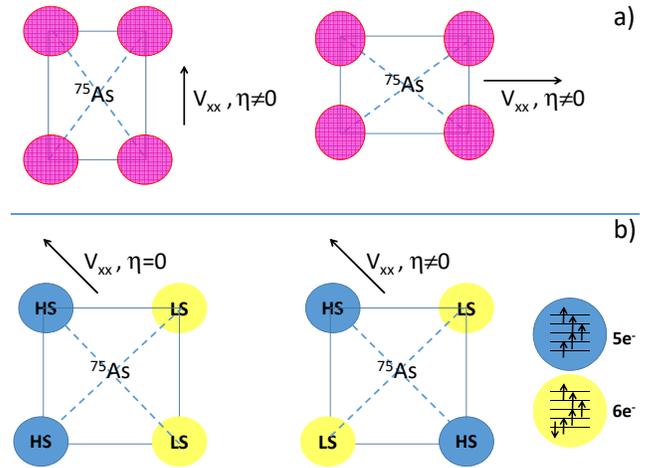} \caption{ (Color online) a) Configuration
of the four nearest neigbour Fe ion of a given $^{75}$As nucleus,
for a homogeneous charge distribution with 5.5 electrons per Fe,
in the presence of a tetragonal to orthorombic distortion induced by the nematic order. The configuration on the
left and on the right are characterized by an EFG with $\eta\neq
0$, with the X principal axis along orthogonal directions, as
depicted by the arrows. Since $\eta$ and $V_{zz}$ are the same for
both configurations, according to Eq. \ref{nqrspe} the NQR frequency
is the same. b) Configuration of the four nearest neigbour Fe ion
of a given $^{75}$As nucleus, in presence of charge
disproportionation promoting both HS 5 electron and a LS 6
electron configurations on Fe, with no breaking of the rotational symmetry. Both
configurations are characterized by the same  $V_{zz}$ but by
$\eta=0$ (left) and by $\eta\simeq 0.28$ (right, derived from a point charge model) which, from Eq.
\ref{nqrspe}, lead to a change of the NQR frequency by about 150
kHz. } \label{FigEFG}
\end{figure}

On the basis of $^{75}$As NMR measurements Li et al.\cite{LiNMR}
have pointed out that the broadening of the NMR spectra for $T< T^*$ could be due to the onset of a nematic order breaking
the rotational symmetry.\cite{Kasahara}
In the NMR experiments this gives rise to different possible
orientations of the external magnetic field with respect to the EFG tensor
principal axis (see Fig.\ref{FigEFG}a) and accordingly to a
broadening or a splitting of the NMR spectra.\cite{LiNMR,Iye} However,
in NQR, where there is no applied magnetic field, the breaking of the tetragonal symmetry should not give 
rise to a broadening or a splitting of the spectra. It causes a drop of $V_{zz}$ \cite{LaFeAsO} and an increase of the asymmetry parameter $\eta$ at $T_{nem}$ \cite{Kitagawa,Ok,Zheng}, and the different signs of the change of $V_{zz}$ and of $\eta$ counteract in Eq. \ref{nqrspe}, leading at most to a slight shift in the NQR peak frequency. 

One can expect a formation of domains characterized by a rotation of the X and Y axes, as is depicted in Fig. 6a. This formation of domains is characteristic for all iron pnictides and chalcogenides \cite{Kitagawa,Ok,Zheng,BaekOrb,LaFeAsO}, however, it does not affect $V_{zz}$ nor $\eta$ and hence, according to Eq. \ref{nqrspe}, just one NQR line should be detected. In fact, in BaFe$_2$As$_2$ and in LaFeAsO, where a $B_{2g}$ nematic order develops, there is no evidence for a distribution of quadrupolar frequencies and the zero-field $^{75}$As NMR lines remain rather narrow down to low temperature.\cite{LaFeAsO,LaMn} Hence, the nematic order cannot justify the broadening or splitting of the NQR spectra starting at T$^*$. 

On the other hand, in \rbfe crystals a nematic order has been recently detected by elastoresitivity measurements at $T_{nem}\simeq 38$ K$\ll T^*$.\cite{XYnematic} This nematic order has a $B_{1g}$ symmetry resulting in a divergence of the nematic susceptibility when the strain is applied along the [100] direction. Around T$_{nem}$ a critical slowing down of the nematic fluctuations is expected to enhance the low-frequency spectral density probed by $1/T_1$. Remarkably, in \rbfe crystal we observe a clear anomaly in the temperature dependence of $1/T_1$ around 35 K (Fig.\ref{FigT1nem}), very close to where elastoresistivity measurements detect a divergence in the nematic susceptibility.\cite{XYnematic} It is noticed that this anomaly is absent in the NMR experiments on the same crystal, which suggests either an effect of the external magnetic field on the nematic fluctuations or a marked frequency dependence of the fluctuations. By subtracting from the raw data the power law behaviour describing the contribution of the correlated spin dynamics to $1/T_1$ for $T< T^*$, we derived $1/T_{1nem}$, the contribution of nematic fluctuations to $^{75}$As spin-lattice relaxation rate (Fig. \ref{FigT1nem}). A clear peak in $1/T_{1nem}$ is detected at $T_{nem}$, as expected for the critical slowing down of the fluctuations, which causes in two dimensions a divergence of $1/T_{1nem}\propto (T-T_{nem})^{-z\nu}$, with $z$ a dynamic scaling exponent and $\nu$ the correlation length critical exponent.\cite{critical,carretta1997} Given the error bars in $1/T_1$ data we cannot draw any clear cut conclusion on the critical exponents, still we observe that the data are consistent with a mean-field exponent $\nu=0.5$ and $z=1$.\cite{criticaldue} 

Hence, although from NQR $1/T_1$ measurements alone one cannot conclude what is generating the peak in $1/T_{1nem}$, the observation of a peak in the nematic susceptibility in the same temperature range suggests that it is probably  driven by electronic nematic fluctuations. We remark that in  BaFe$_2$As$_2$ and in LaFeAsO pnictides, owing to the proximity to the magnetic order, it is not straightforward to separate the contribution of nematic fluctuations to $1/T_1$ from the one of the correlated spin dynamics. In FeSe, where there is no magnetic order, no anomaly in $^{77}$Se $1/T_1$ has been reported at T$_{nem}$.\cite{FeSeNMR} In \csfe we have no evidence of such an anomaly in the spin-lattice relaxation data. This is consistent with elastoresitivity findings\cite{XYnematic} which suggest that T$_{nem}\rightarrow 0$ in this compound. It is important to notice, as it has been pointed out in Ref.\onlinecite{XYnematic},  that XY nematic fluctuations developing at higher temperature could also promote a charge order and explain the observed broadening of the NQR spectra.

A more sound explanation for the NQR line broadening is based on the
realization of a charge disproportionation across the lattice, 
recently proposed by Isidori et al.\cite{Isidori}. The Hund's
coupling $J$ favours the electron transfer among the Fe
ions, in order to realize high spin (HS) configurations, which competes with the on-site Coulomb repulsion $U$. When $J/U$ is sufficiently large configurations with coexisting HS and low spin (LS)
Fe would be favoured, leading to a modulation of the local charge
on the four Fe around the $^{75}$As nuclei.\cite{Isidori} Since more than one
configuration with charge disproportionation can be realized one
can expect a distribution of EFG and accordingly a broadening of
the NQR line, as we observe, as well of the NMR lines, as detected by Li et
al..\cite{LiNMR}

To better  illustrate the effect of charge disproportionation we consider its effect on the basis of a simple point charge model on a cluster formed by 4 Fe ions. 
In Fig.\ref{FigEFG}b we show two possible arrangements of the charge
distribution in the four Fe ions around the $^{75}$As nuclei. Based on the point charge model, it is easy to show that the
two charge configurations are characterized by the same value for
$V_{zz}$ but by quite different values for the EFG asymmetry
$\eta$. Accordingly, from Eq. \ref{nqrspe} the NQR frequency for
the two configurations would differ by a few percent, namely by
few hundreds of kHz, which is of the order of magnitude of the observed
broadening or splitting of the NQR lines in \rbfe and \csfe.

Finally we turn to the discussion of the $\mu$SR results. As it was
pointed out above, close to $T^*$ the $\mu$SR relaxation rate
starts to grow. If the growth of $\lambda$ was not at all related
to the modification of the electronic structure taking place at
$T^*$, one would expect to find a similar behaviour both for \rbfe
and for \csfe, at variance with the experimental findings. The
significant increase in $\lambda$ around $T^*$ suggests that the
changes in the electronic configuration affect the amplitude
and/or the frequency of the fluctuations of the dipolar field
generated by the nuclear spins at the muon site. The fact that the
$\beta_1$ exponent characterizing the decay of the muon
polarization is varying between about 1 (exponential decay) for
$T> T^*$ to values approaching 2 (Gaussian decay) for $T<T^*$,
suggests a slowing down of the fluctuations at $T^*$,\cite{muon} a situation
typically observed when muons stop diffusing through the lattice.
Hence, we argue that the growth of $\lambda$ for $T<T^*$ indicates
a slowing down of the muon diffusion, induced by the changes in
the lattice potential taking place around $T^*$.

\section{\label{sec:conclusion} Conclusions}
In conclusion, we have shown that in \rbfe and \csfe superconductors a reorganization of the electronic structure takes place at a temperature $T^*$, below which not only a change in the spin-lattice relaxation  but also an increase in the linewidth of $^{75}$As NQR spectra are detected. That broadening suggests the presence of a inhomogeneous charge distribution in the lattice which could be driven by the competition between Hund's coupling and the Hubbard repulsion. The changes in the lattice potential at $T^*$ possibly affect also the muon diffusion and explain the growth in the muon relaxation detected around that temperature. At T$_{nem}\ll$T$^*$ a nematic order develops in \rbfe and the critical slowing down of the fluctuations can naturally explain the peak observed in $1/T_1$. These results point out for the coexistence of electronic nematic and charge orders in \rbfe, with a mechanism driving the charge modulation different from that driving the CDW phase in the cuprates. 

\section*{acknowledgments}
The authors wish to thank L.Fanfarillo and M.Capone for useful discussions. P. Biswas and H.Luetkens are acknowledged for their help during the measurements at ISIS and at PSI, respectively. The authors thank  S. Müller-
Litvanyi, for support with SEM-EDX measurements.  SA thanks the DFG for funding (AS 523/4-1 and 523/3-1). The work at IFW was supported by the Deutsche Forschungsgemeinschaft (DFG) through the Priority Programme SPP1458.
The activity in Pavia was supported by  MIUR-PRIN2015 Project No. 2015C5SEJJ.


\begin{thebibliography}{99}
\bibitem{Tstar} There is an uncertainty of $\pm 5$ K in the estimate of $T^*$ either from the temperature at which the linewidth starts to increase montonously down to low temperatures or from the temperature at which $1/T_1$ crosses from being nearly constant to a power law behaviour.
\end{thebibliography}
\end{document}